\documentclass[12pt,preprint]{aastex}         
\usepackage{psfig}
\def\be{\begin{equation}}
\def\ee{\end{equation}}

\def\lsim{\lower 2pt \hbox{$\, \buildrel {\scriptstyle <}\over
         {\scriptstyle \sim}\,$}}
\newcommand\gsim{\buildrel > \over \sim}
\begin{document}
\newcommand{\figureout}[3]{\psfig{figure=#1,width=5.5in,angle=#2} 
   \figcaption{#3} }

\title{High-Altitude Particle Acceleration and\\
Radiation in Pulsar Slot Gaps}

\author{Alex G. Muslimov\altaffilmark{1} \& 
Alice K. Harding\altaffilmark{2}}   

\altaffiltext{1}{ManTech International Corporation, 
Lexington Park, MD 20653}

\altaffiltext{2}{Laboratory of High Energy Astrophysics,      
NASA/Goddard Space Flight Center, Greenbelt, MD 20771}
 

\begin{abstract}
We explore the pulsar slot gap (SG) electrodynamics up to very high altitudes, where for most relatively 
rapidly rotating pulsars both the standard 
small-angle approximation and the assumption that the magnetic field lines are ideal stream lines break 
down. We address the 
importance of the electrodynamic conditions at the SG boundaries and the occurrence of a steady-state 
drift of charged particles across the SG field lines at very high altitudes. These boundary conditions and the 
deviation of particle trajectories from stream lines determine the asymptotic behavior of the scalar potential at all radii from the 
polar cap (PC) to
near the light cylinder. 
As a result, we demonstrate that the steady-state accelerating electric field, $E_{\parallel}$, must approach a small 
and constant value at high altitude above the PC.  This $E_{\parallel }$ is capable of maintaining electrons moving with high Lorentz factors ($\sim {\rm a~few}~ \times 10^7$) and emitting curvature $\gamma $-ray photons up to nearly the light cylinder.  By numerical simulations, we show that primary electrons accelerating from the PC surface to high altitude in the SG along the outer edge of the open field region will form caustic emission patterns on the trailing dipole field lines.  Acceleration and emission in such an extended SG may form the physical basis of a model that can successfully 
reproduce some pulsar high-energy light curves.
\end{abstract} 


\pagebreak
  
\section{INTRODUCTION}
There is no doubt that pulsars are accelerating particles up to relativistic energies in their magnetospheres, 
and that these particles are primarily responsible for the pulsar radio- to high-energy non-thermal emission. It 
is also believed that the energetics of this acceleration, as well as the main physical processes involved in production 
of high-energy photons, are more or less understood. However, the ambiguity in interpretation of pulsar timing 
observations in terms of emission site mapping in a pulsar magnetosphere makes it difficult to answer the basic 
question of where the pulsar high-energy emission originates.  In their recent attempt to explain the observed 
high-energy light curves of pulsars, Dyks \& Rudak (2003) concentrated on a purely geometrical 
model by postulating that 
the emission is produced in a relatively narrow region along the last open magnetic field lines of a pulsar 
magnetosphere. The interesting result of their study is the occurrence of caustic emission zones (Morini 1983), i.e. the phase shifts of radiation emitted at radii between $\sim 0.1 - 0.7$ times the light cylinder radius, parallel to field lines on the trailing edge of the polar cap (PC),
are cancelled by phase shifts due to relativistic effects of aberration and time-of-flight.  Radiation emitted
over a large range of altitudes thus arrives in phase, forming two narrow peaks in the light curves, very similar to those of known $\gamma$-ray pulsars (e.g. Thompson 2001).

In our previous paper (Muslimov \& Harding 2003 [MH03]) we began discussing the regime of acceleration 
of particles and production of high-energy emission within the pulsar slot gap (SG), a narrow region on the 
boundary of the open field lines, where the electric field drops to zero.  The SG is a pair-free region of 
slower acceleration, in which the parallel electric field is unscreened.  
Pair cascades develop along the inner edge of the SG at several stellar radii above the NS surface.  
Even though the SG regime in pulsars was originally introduced 
in the electrodynamic model of Arons \& Scharlemann (1979), it was not considered a viable high-energy emission
region (see e.g. Arons 1996). The revised version of the SG regime proposed by MH03 incorporates the effect of 
relativistic frame dragging (Muslimov \& Tsygan 1992 [MT92]) and, more importantly, the effect of SG boundaries on the strength of the accelerating 
electric field within the SG.  MH03 demonstrated that the primary electrons tend to accelerate up to higher 
altitudes before pair production begins, and pair cascades continue along the inner boundary of the SG 
until the magnetic field becomes too low.  The resulting radiation from the pair cascades forms a wide, hollow
cone of high-energy radiation due to the flaring of field lines.  Adhering to the small-angle approximation, 
MH03 restricted their study of the SG regime to 
altitudes less than four-five stellar radii.  However, since the parallel electric field in the SG is not screened
on field lines close to the open-field boundary, acceleration may continue to much higher altitudes.  Particle
acceleration and radiation in such an extended SG may therefore provide a physical basis for the two-pole caustic
model of Dyks \& Rudak (2003).

Formation of a SG requires the production of enough pair multiplicity to screen the parallel electric field 
above the pair formation front.  We have found from our previous studies (Harding \& Muslimov 2001 [HM01], 
2002 [HM02]) 
that the youngest and most energetic
pulsars can produce pairs from curvature radiation (CR) of primary electrons, which are numerous enough to screen the electric field.  Older, less energetic pulsars, those below the CR pair death line, can 
produce only pairs from inverse Compton radiation of primary electrons scattering thermal X-rays from the NS surface.  The inverse Compton pairs are not numerous enough to completely screen the parallel electric field.  A necessary
condition for formation of a SG is thus the ability to produce pairs from CR, and the expression for the CR death
line (given by Eqn [52] of HM02) defines the boundary in the $P$-$\dot P$ diagram of pulsars capable of having SGs.
Such pulsars include the Crab, Vela, Geminga and most of the $\gamma$-ray pulsars detected by EGRET, but not the majority of millisecond pulsars.

The extension of the regime of SG acceleration to much higher altitudes is the main subject of the present paper. 
In the Sections below we outline our approach to constructing an appropriate steady-state physical 
solution that can be used up to very high altitudes in the SG. We also discuss the immediate consequences of our 
proposed extended SG solution: acceleration of particles (electrons, positrons) and high-energy emission up to 
nearly light-cylinder radius, and the possibility of occurrence of high-altitude caustic emission on trailing 
field lines.    

The paper is organized as follows. In \S 2 we discuss the electrodynamics 
within the SG regions of pulsars. We address the physical constraints on the scalar potential (\S 2.1) and  
equipotentiality of SG boundaries and derivation of effective Poisson's equation (\S 2.2) in the outermost 
section of SG. In \S 3 we present the electrodynamic solution within the SG at very high altitudes. In \S 3.1 
we illustrate the possibility of extended acceleration within SG in the regime where the acceleration is 
balanced by the curvature-radiation reaction. Our numerical calculations are discussed in \S 3.2. Finally, 
in \S 4 we discuss our main results and draw our principal conclusions.

\section{Steady State SG Electrodynamics}

In the frame of reference rigidly corotating with a neutron star (NS), where the 
magnetic field is stationary (and having pure dipolar geometry), the general relativistic Maxwell's equations 
yield (see MT92)
\be
{\bf E} - {1\over \alpha c} ({\bf w}-{\bf u})\times {\bf B} = 
- {1\over \alpha }{\bf \nabla } \Phi ,
\label{gradPhi}
\ee
where {\bf E} and {\bf B} are the electric and magnetic fields defined in 
Zero-Angular-Momentum-Observer (ZAMO) frame of reference (see Macdonald \& Thorne 1982), {\bf u} 
is the rotational velocity, and {\bf w} is the differential velocity of rotation of inertial frame of 
reference, $\Phi $ is the scalar potential, and the so-called geneneral-relativistic `lapse function', 
$\alpha $, is defined below, right after expression (\ref{F}). Taking the divergence of 
eq. (\ref{gradPhi}) and making use of Maxwell equation 
\be
{\bf \nabla } \cdot {\bf E} = 4 \pi \rho ,
\label{divE}
\ee 
we get the Poisson's equation for the scalar potential $\Phi $
\be
{\bf \nabla }\cdot \left( {1\over \alpha} \nabla \Phi \right) = 
4 \pi (\rho_{_{ \rm GJ}} - \rho),
\label{Poisson}
\ee
where 
\be
\rho _{_{\rm GJ}} = - {1\over {4 \pi c}} {\bf \nabla }
\cdot \left[ {1\over \alpha } ({\bf u}-{\bf w}) \times 
{\bf B} \right] 
\label{rhoGJ}
\ee
is the general relativistic expression for the Goldreich-Julian (GJ) charge density (cf. Goldreich \& Julian, 1969), 
and $\rho $ is the actual charge density of electrons determined by their relativistic flow along the magnetic field 
lines and which is fixed by the condition $E_{\parallel } = 0$ at the stellar surface (see MT92 for details). 
Thus our electrodynamic description of charges streaming along the open field lines will imply the 
space-charge-limited flow approximation (at least near the stellar surface, within the radial distance of 
less than a few stellar radii).  

Note that the l.h.s. of eq. (\ref{gradPhi}) can be treated as the effective electric field in the frame 
of reference rigidly corotating with the NS,   
\be
{\bf E'} = {\bf E} - {1\over {\alpha c}} ( {\bf w} - 
{\bf u} ) \times {\bf B} .
\label{E'}
\ee
The condition of absence of any electric field in the regions of the magnetosphere with closed field lines 
rigidly corotating with the NS, ${\bf E'}= 0$, necessarily implies that these regions should be 
filled with charges of density $\rho = \rho _{_{\rm GJ}}$ (as a trivial solution of eq. [\ref{Poisson}]).

Our previously derived solutions (see MT92, Muslimov \& Harding 1997 [MH97]) for the case 
$|\rho| \lsim |\rho _{_{\rm GJ}}|$ were limited 
by a small-angle approximation and therefore cannot be justifiably used beyond the radial distances of about 
$\sim $ 3-4 stellar radii above the PC surface (for a typical pulsar spin period). Recently, MH03 discussed 
the SG solution which is also formally limited 
to a small-angle approximation. This means that, because of the curving of the SG toward the magnetic equator, the 
solution derived in MH03 cannot be used for high altitudes, typically exceeding a few (or several, 
at most) stellar radii above the surface. Here we discuss the regime of steady-state acceleration 
within the SG extending up to very high altitudes, nearly approaching the 
light-cylinder. By addressing the basic physical conditions that are required for the occurrence of this 
regime, we observe that the standard concept of GJ charge density becomes inapplicable in the outermost section 
of the SG, where the effective GJ charge density gets significantly constrained by the requirement of equipotentiality 
of SG boundaries and by the effect of cross-field motion of charges. 

We shall explore the SG solution in the outermost part of NS magnetosphere (but still within the light-cylinder) 
satisfying the same boundary conditions as those used in all our previous studies: equipotentiality of the SG 
surfaces, and zero-electric field condition at the PC surface. We propose the following {\it ansatz} for constructing the 
general solution extending from the innermost section up through the outermost section of the SG. By considering the property (see Section 2.1 below) of the scalar potential $\Phi $ at large (up to the light-cylinder radius) distances  found by 
Mestel et al. (1985) together with the abovementioned boundary conditions 
(equipotentiality of SG surfaces and 
zero-electric field condition at the PC surface), we can unambiguosly constrain and determine the outermost 
solution for the potential $\Phi $. Then, by matching the outermost solution with the known innermost solution near 
the NS surface (presented in MH03) we can construct the approximate general solution applicable to both the innermost and outermost sections of the SG. 

Note that any physically meaningful electrodynamic solution in the outermost part of the magnetosphere should 
take into account the effect of particle drift across the field lines or deviating of particle trajectories 
from magnetic field lines. This effect should unavoidably constrain the scalar potential $\Phi $, simply because in this 
region the field lines cannot be treated as characteristics or as stream lines for the flux of relativistically 
moving electrons. In a steady-state situation, it is reasonable to expect that the scalar potential $\Phi $ 
is a monotonically increasing (or decreasing, as in the case of acceleration of positive charges) and then saturating function of radial distance so that the outermost solution 
gradually matches the innermost one. In this study we demonstrate that in a steady-state situation, the 
constraint on $\Phi $ in the outermost region of the NS magnetosphere (but well within the light cylinder) along 
with the condition of equipotentiality of the SG surface allows us to derive an appropriate  electrostatic 
solution. This solution implies initial (in the innermost section of a SG) boosting of electron acceleration 
over characteristic lenghtscale of $\sim 1-2$ stellar radii and subsequent extremely slow post-boost acceleration  
over lengthscale $\sim $ light-cylinder radius.    

\subsection{Constraint on Potential $\Phi $ in the Outermost Section of SG}

The equation of motion of an electron of mass $m$ and charge -$e$ can be written as (see eq. [2.24] 
in Mestel et al. 1985) 
\be  
-{e\over m} \left( {\bf E} + {{\bf v}\over c}\times {\bf B} \right) 
= {\bf v}\cdot {\bf \nabla }(\gamma {\bf v}) = 
{\bf \nabla } (\gamma c^2) - {\bf v}\times ({\bf \nabla }\times 
(\gamma {\bf v})), 
\label{eqnmotion}
\ee
where $\gamma = (1-v^2/c^2)^{-1/2}$, ${\bf v}$ is the electron velocity. In Mestel et al.'s notations, 
\be
{\bf E} = - \nabla \phi,
\label{phi_Mestel}
\ee
and 
\be
{{\bf v}\over c}\times \left({\bf B}-{{mc}\over e} {\bf \nabla} 
\times (\gamma {\bf v})\right)= {\bf \nabla} \phi ^{\ast },
\label{phi*_Mestel}
\ee
so that equation of motion, (\ref{eqnmotion}), translates into (see eq. [2.26] in Mestel et al. 1985) 
\be
-e\phi + \gamma mc^2 = -e\phi ^{\ast }(S),
\label{phi-phi*}
\ee      
where $S$ is the stream function. 

Note that the scalar potential $\Phi $ defined by eq. (\ref{gradPhi}) is a general-relativistic counterpart 
of the so-called `non-corotational' potential $\psi $ (see e.g. eq. [2.5] in Mestel et al. 1985) introduced 
by Endean (1974), Mestel (1973), and Westfold (1981) and which can be defined via equation
\be
\nabla \psi = \nabla \phi - {{{\bf \Omega }\times {\bf r}}\over c}\times {\bf B} .
\ee
For relatively low altitudes $\Phi \approx \gamma (mc^2/e)$. For high altitudes (but 
still within the light-cylinder), as was first demonstrated by Mestel et al. (1985), the change in the 
angular momentum $\gamma m \Omega {\tilde r}^2$ (where ${\tilde r}$ is the radial cylindrical polar coordinate) of a 
streaming particle occurs only through the toroidal component of the magnetic force, requiring departure 
from strict flow along the field lines. The combination of energy and angular momentum integrals gives 
(in cylindrical polar coordinates in the axisymmetric case)
\be
\psi = {{mc^2}\over e} \gamma \left( 1 
- {{\Omega ^2 {\tilde r}^2}\over c^2}\right) - \Gamma (S),
\label{psi}
\ee
where $\Gamma (S)$ is some function which is constant on stream lines and which is set at the stellar 
surface. Here the term $\propto \Omega ^2$ is the so-called `centrifugal-slingshot' term (see Mestel 
et al. 1985) arising from the flow of electrons across the field lines. 

\noindent
It is important to point out that within the domain of the SG the actual deviation from strict flow 
along (poloidal) $\bf B$ is of order 
\be
\delta \approx \gamma mc^2/(eB\Delta l_{_{\rm SG}}),
\label{delta}
\ee
where 
$\Delta l _{_{\rm SG}}\sim \theta _0 R \Delta \xi _{_{\rm SG}}\sqrt{\eta _{\ast }}$ is the characteristic 
latitudinal SG thickness at dimensionless radial distance $\eta _{\ast }$ ($\eta =r/R$); $\theta _0$ is the 
PC half-angle, 
$\theta _0 = [\Omega R/f(1)c]^{1/2}$; and $\Delta \xi _{_{\rm SG}}$ is the latitudinal SG thickness 
in units of $\xi $($\xi = \theta /\theta _0$ is the dimensionless colatitude of a PC field; see Section 3 
and also MH03 for details). Thus, the characteristic dimensionless radial distance at which the magnitude 
of the deviation from strict flow along (poloidal) $\bf B$ reaches $\delta $, can be estimated as 
\be
\eta _{\ast }~\sim 0.6~\eta _{lc}~(\Delta \xi _{_{\rm SG}}~\delta )^{2/5}~{R_6\over P_{0.1}}
\left( {{R_6^3B_{12}^2}\over{\gamma _7^2P_{0.1}}} \right) ^{1/5},
\label{sling-shot}
\ee
where $\eta _{lc} = c/\Omega R$ is the dimensionless radius of the light-cylinder; $\Omega = 2\pi/P$ is the 
angular velocity of NS rotation; $R$ and $P$ are the NS radius and spin period, respectively; 
$B_{12}=B_0/10^{12}$ G, $B_0$ is the surface value of NS magnetic field strength; $R_6=R/10^6$ cm, $P_{0.1}=P/0.1$ s, 
and $\gamma_7 = \gamma /10^7$.

\noindent 
For the parameters of the Crab pulsar (here we adopt the values: $B_{12}=8$, $R_6=1.6$, and $P_{0.1}=0.33$) and assuming $\delta \sim 
0.05-0.1$, from formula (\ref{sling-shot}) we get
\be
\eta _{\ast } = (0.3-0.4)~\eta _{lc}~\left( {{\Delta \xi _{\rm SG}}\over {0.01}} {3\over {\gamma _7}} \right) ^{2/5}, 
\label{eta_star}
\ee
so that for the estimated value of  $\Delta \xi _{_{\rm SG}} \sim 0.05$ (see MH03) and for $\gamma _7 \sim 3$ 
condition (\ref{sling-shot}) is satisfied at $\eta _{\ast }\sim (0.5-0.7)~\eta _{lc}$.

\noindent 
Formula (\ref{eta_star}) means that within the SG the effect of transfield motion 
becomes important already at $\eta \lsim \eta _{lc}$ (for the Crab-like pulsars) and should be taken into account. This is 
significantly different from Mestel et al.'s model where a similar situation would occur well beyond 
the light cylinder (and where $\Delta l_{_{\rm SG}}$ in formula [\ref{delta}] should be replaced by the 
light cylinder radius, $R_{lc} = c /\Omega$), because the lengthscale was the entire open field region 
rather than the narrow SG.   
   
The fundamental consequence of eq. (\ref{psi}) is that in the region where the centrifugal-slingshot 
effect becomes important (e.g. at $\eta \lsim \eta _{lc}$ for the Crab-like pulsars) the electrons 
crossing the field lines begin picking up energy from the corotational 
part of the potential (from potential $\phi $, in Mestel et al.'s notation), so that in the regime of mostly 
transverse flow the change in potential (change in $\Gamma (S)$) across the field lines caused by rotation 
tends to balance the change in the corotational part of the potential. Suppose that the 
electrons are flowing with relativistic velocities along the magnetic field lines and entering the 
region where they are getting `decoupled' from the magnetic field lines. Apparently, the solution of 
MH03 for $\rho $ 
and therefore for $\Phi $ will not be warranted in this region and especially in the region with 
predominantly transfield flow. However, we may justifiably assume that at the onset of the transfield 
flow regime, where the relativistic flow of electrons is still mostly along the magnetic field lines, the 
condition (see e.g. Mestel 1995, 1999) 
\be
{\bf E'}_{\perp} \approx {\bf E}_{\perp } + {{{\bf \Omega } \times {\bf r}}\over c} 
\times {\bf B} \approx 0,
\label{Eperp}
\ee
`turns on', with $E_{\parallel } \ll E_{\perp }$ (for most acceleration scenarios we discuss in this paper 
$E_{\parallel }$ is balanced by the CR reaction force).  Further out in the magnetosphere the condition (\ref{Eperp}) 
may transform into the perfect MHD condition (see also Contopoulos, Kazanas \& Fendt, 1999), 
${\bf E}+({\bf v}/c)\times {\bf B} = 0$, which is a good approximation as long as $E_{\parallel } \ll E_{\perp }$.

We suggest that the condition (\ref{Eperp}) together with the  equipotentiality of the SG surfaces may in fact 
determine the behavior of potential $\Phi $ through the outermost section of the SG. It is important that 
at large radial distances ${\bf E'}_{\perp }$ as given by 
eq. (\ref{E'}) is dominated by the term $({\bf \Omega }\times {\bf r}/c)\times {\bf B}$, 
which tends to produce significant electric potential drop across the SG, and as a result, would induce enormous 
surface charge on the SG boundaries. Also, it must be pointed out, that this term would tend to induce a strong 
component of $E_{\parallel }$ with the polarity that may change with altitude and become opposite to that of 
the main component of $E_{\parallel }$, 
produced by a small imbalance between the GJ and actual charge densities in the innermost region, at altitudes 
within $\sim 1-2$ stellar radii. The occurrence of a strong component of $E_{\parallel }$ with reversed polarity 
would unavoidably disrupt the continuous flow of electrons along the field lines and result in an essentially  
non-stationary regime of particle flow. However, in the steady-state situation we consider in this paper, the 
occurrence of cross-field motion of electrons would effectively screen out the excessive GJ space charge and 
short out the SG boundaries, thus maintaining them as equipotential. The latter means that 
condition (\ref{Eperp}) would be roughly satisfied in and beyond this region. Before we discuss 
how condition (\ref{Eperp}) can be explicitly incorporated into our electrostatic solution, let us discuss 
the consistency of this condition with the assumption of equipotentiality of SG boundaries all the way from 
the PC surface up to the very high altitudes (say, up to $\sim 0.1-0.5$ of the light-cylinder radius).

\subsection{Equipotentiality of SG Boundaries and Effective Poisson's Equation in the Outermost Section of SG}

Let us consider the cross-sectional area of a magnetic flux tube (of dipole field) emanating from the PC at 
radial distance $\eta $ ($= r/R$)
\be
{\cal S}(\eta ) = {\cal S}(1) {{f(1)}\over {f(\eta )}} \eta ^3 ,
\label{S}  
\ee
where $f(\eta )$ is the general-relativistic correction factor, defined as (see e.g. MT92)
\be
f(\eta )  = -3 \left( {{\eta }\over {\varepsilon }} \right) ^3 \left[ \ln \left( 1 - 
{{\varepsilon }\over {\eta }}\right) 
+ {{\varepsilon }\over {\eta }} \left( 1 + {{\varepsilon }\over {2\eta }}\right) \right],
\label{f()} 
\ee
where $\eta = r/R$ is the dimensionless radial coordinate, $\varepsilon = r_g/R$, and $r_g$ is the gravitational radius 
of the NS.  One can use approximate formula $f(\eta ) \approx 1+0.75x+0.6x^2$ (where $x=\varepsilon /\eta $). 
For a canonical NS of 1.4 solar mass and 10 km radius ($\varepsilon = 0.4$) $f(1) \approx 1.4$. 

\noindent
In this paper, as in our previous studies, we use the magnetic spherical polar coordinates 
($\eta, \theta , \phi _{pc}$).  We also denote by $\chi$ the pulsar obliquity (angle between the NS rotation axis and 
magnetic moment). We will refer a `normal polarity' pulsar as one having $0^{\circ} \leq \chi < 90^{\circ}$ 
(north magnetic pole near north astrographic pole: ${\bf \Omega} \cdot {\bf m} > 0$, where $\bf m $ is 
the NS magnetic dipole moment), and a `reversed polarity' pulsar as one having $90^{\circ} < \chi \leq 180^{\circ}$ 
(north magnetic pole near south astrographic pole: ${\bf \Omega }\cdot {\bf m } < 0$). 

\noindent 
We shall now introduce the flux of charges streaming with relativistic velocity through ${\cal S}(\eta )$, 
\be
F = \alpha (\eta )c {\cal S}(\eta ) \rho (\eta ),
\label{F}
\ee 
where $\rho $ is the local charge density, and $\alpha = (1-\varepsilon/\eta )^{1/2}$ is the lapse function. 

\noindent
In a steady-state regime, well within the light cylinder, the flux $F$ should be constant along the 
individual magnetic flux tube as a consequence of charge continuity equation which implies that  
$\alpha \rho \propto B \propto \eta ^{-3}$. In this case $F$ is a function of $\xi $ only (and not 
$\eta $). [The variable $\xi $ is equivalent to the Stokes stream function $S$ (see eq. [\ref{phi-phi*}]) 
used by Mestel et al.] However, at very high altitudes, for any stream line within the SG we can write 
that 
\be
B = {{B_0}\over f(1)}{\beta \over {\eta ^3}} , 
\label{B}
\ee
where $\beta = \sqrt{1-3\eta /4\eta _{lc}}$ is a factor that takes into account the change in the geometry 
of the flux tube as we move from the magnetic pole to the equator.

The explicit expressions (see e.g. MH97, for the derivation of these expressions) for the GJ charge 
density, $\rho _{_{\rm GJ}}$, and actual charge density, $\rho $, may be written as  
\be
\rho _{_{\rm GJ}} = - \rho _0 {{f(\eta )}\over {f(1)}}
{1\over {\alpha \eta ^3}} \left\{ [ a_0 (\xi ) + a_1(\eta , \xi) ] 
\cos \chi + [ b_0(\xi ) + b_1(\eta, \xi )] \sin \chi \cos \phi _{\rm pc}\right\},
\label{rhoGJ_2}
\ee
and 
\be
\rho \approx - \rho _0 {{f(\eta )}\over {f(1)}}
{1\over {\alpha \eta ^3}} \beta (\eta )\left\{ [ a_0 (\xi ) + a_1(1, \xi) ] 
\cos \chi + [ b_0(\xi ) + b_1(1, \xi )] \sin \chi \cos \phi _{\rm pc}\right\},
\label{rho}
\ee
respectively (see also MT92 for the exact expression for $\rho $ valid for arbitrary small altitudes and 
which is consistent with the radial profiles of potential $\Phi $ and $E_{\parallel }$). 
Here $\rho _0 = \Omega B_0 /2\pi c$, and 
\be
a_0 = 1,~~~~~~~~~a_1 = - {{\kappa }\over {\eta ^3}} - {3\over 2} 
H(\eta ) \sin ^2 \theta ,
\label{a0a1}
\ee
\be
b_0 = 0,~~~~~~~~~b_1 = {3\over 2} H(\eta ) \sin \theta \cos \theta ,
\label{b0b1}
\ee
\be
H(\eta) = {{\varepsilon }\over {\eta }} - {{\kappa }\over {\eta ^3}} + 
{1\over {(1-\varepsilon /\eta) f(\eta)}}\left( 
1-{3\over 2} {\varepsilon \over \eta } + {\kappa \over 2\eta ^3} 
\right).
\label{H}
\ee
Here $\kappa $ is the general-relativistic parameter characterizing the magnitude of the frame-dragging effect near 
the stellar surface measured in stellar rotation velocity, $\Omega $, and $\beta $ is defined right after 
eq. (\ref{B}).  These expressions are formally derived for 
arbitrarily large distances but still within the light cylinder. Well 
within the light cylinder, at $\eta \ll \eta _{lc}$ ($\eta _{lc} \sim \theta _0^{-2}$), and in a small-angle 
approximation, 
\be
a_1 \approx - {\kappa \over \eta ^3},~~~~~~~~
b_1 \approx {3\over 2} 
\theta _0 H(\eta ) \sqrt{\eta {f(1)\over f(\eta )}} , 
\label{a1b1}
\ee
which correspond to the expressions for $\rho $ and $\rho _{_{\rm GJ}}$ we used in our previous papers for the 
situations where a small-angle approximation was in fact more than satisfactory. One can also use the approximate 
expression, $H(\eta ) \approx 1-0.25x-0.16x^2-0.5(\kappa/\varepsilon ^3)x^3(1-0.25x-0.21x^2)$, where 
$x= \varepsilon /\eta $. For a canonical NS [see also eq. (\ref{f()})] $H(1) \approx 0.8$.   

Now, the flux of effective GJ charges can be written as a sum of constant and varying with altitude 
components,
\be
F_{_{\rm GJ}} = F_{_{\rm GJ,0}} + F_{_{\rm GJ,1}} ,
\label{FGJ}
\ee  
where 
\be
F_{_{{\rm GJ},{\rm i}}} = F_0 ( a_i \cos \chi + b_i \sin \chi \cos \phi _{\rm pc} ),~~~~~i = 0,~1~~,
\label{FG1}
\ee
and $F_0 = - \rho _0 c {\cal S}(1)$ is a constant factor depending on pulsar bulk parameters.
Note that, according to the GJ reasoning (Goldreich \& Julian, 1969), the corotating region of the pulsar 
magnetosphere with closed magnetic field lines should be filled with charges of local density 
$\rho _{_{\rm GJ}}$. This condition guarantees that any electric field which could be possibly generated 
in this region should be completely screened out. We do not intend to speculate on the dynamics of formation 
of the non-vacuum pulsar magnetosphere (see e.g. Krause-Polstorff \& Michel, 1985; Arons \& Spitkovsky, 2002), 
but we assume that the filling in of the closed field lines of the initially charge-starved 
magnetosphere with charges is most likely to occur along the field lines. Even more, since $F_{_{\rm GJ,1}}$ is 
a function of $\eta $, one may expect that this process develops in a non-stationary manner.

Let us now consider the flux $\Delta F$, corresponding to the charge imbalance between the GJ and actual 
local space-charge density, 
$\Delta \rho = \rho _{_{\rm GJ}}-\rho $, 
\be
\Delta F = F_{_{\rm GJ}}-F \approx \alpha c {\cal S}(\eta ) (\rho _{_{\rm GJ}}-\rho ) 
\approx \Delta F_{\ast }(\eta , \xi) - \beta (\eta) \Delta F_{\ast }(1,\xi ), 
\label{DeltaF1}
\ee
where 
\be
\Delta F_{\ast }(\eta, \xi ) = F_0[a_1(\eta , \xi ) \cos \chi 
+ b_1(\eta , \xi ) \sin \chi \cos \phi _{\rm pc}]
\label{DeltaF_ast}
\ee 
Using the explicit expressions for $a_1$ and $b_1$ (see eqs [\ref{a0a1}], [\ref{b0b1}]) we can write 
\begin{eqnarray} 
\Delta F & \approx & F_0 \left\{ \left[ 
\kappa \left( \beta -{1\over \eta ^3}\right) + 1-\beta + {3\over 2} H(1) \theta _0^2 
\left( \beta -\eta {f(1)\over f(\eta )} {{H(\eta )} \over {H(1)}} \right) \right] 
\cos \chi \right. + \nonumber \\
& & \left. {3\over 2} H(1)\theta _0 \left( {H(\eta )\over H(1)} 
\sqrt{\eta {f(1)\over f(\eta )}} -\beta \right) \sin \chi \cos \phi _{\rm pc} \right\},
\label{DeltaF2}      
\end{eqnarray}
where $\beta $ is defined right after eq. (\ref{B}). 
Here we should reiterate that the formal usage of a general expression for 
$\Delta \rho = \rho _{_{\rm GJ}}-\rho $ ($=\Delta F/\alpha c S[\eta ]$) to solve the Poisson's equation 
for arbitrarily large altitudes leads to the inconsistent and even erroneous result, mostly because of the physical 
reasons discussed 
in the end of previous Section. Namely, such a solution would imply the building up of the effective flux, $\Delta F$, exceeding the GJ flux $F_{_{\rm GJ}}(\eta =1)$, fixed at the stellar surface, and sign reversal 
of the accelerating electric field at high altitudes. Also, we would like briefly comment on the space-charge-limited 
flow approximation at high altitudes in pulsars. Generally, the space-charge limitation occurs when the ejected 
charges reduce the accelerating potential drop boosting the initial particle energy. In pulsars, the flux of 
electrons ejected from the PC surface is 
limited by the value of GJ space charge at the bottom of the PC. In this case $F$ remains constant along the magnetic 
stream lines and is determined by $F_{_{\rm GJ}}$ at $\eta = 1$. Above the PC surface the space charge of ejected 
electrons reduces the ``vacuum" potential drop by limiting it to the value determined by a small imbalance between 
the GJ charge density and actual charge density of electrons. This approximation is perfectly valid within a few stellar 
radii above the PC surface of most pulsars, when $|\Delta F(\eta )| \lsim |F_{_{\rm GJ}}(1)|$. However, at high 
enough altitudes, where $|\Delta F(\eta )| > |F_{_{\rm GJ}}(1)|$ the situation is akin to the acceleration of test 
particles in a vacuum-like potential drop. Let us examine this in more detail for the SG by using the above 
expressions for $\Delta F$ and then formulate the derivation of an approximate but physically meaningful solution.    

\noindent 
For low altitudes, $\eta \approx 1 + z$ ($z \ll 1$), we can write  
\be
\Delta F \approx 3 F_0 z \left( \kappa \cos \chi + 
{1\over 4} H(1) \theta _0 \sin \chi \cos \phi _{\rm pc} \right),
\label{DeltaF3}
\ee
and 
\be
F_{_{\rm GJ}} \approx F_0 \left[ (1-\kappa ) \cos \chi + 
{3\over 2} H(1) \theta _0 \sin \chi \cos \phi _{\rm pc} \right] , 
\label{FGJ2}
\ee
so that $|\Delta F| \ll |F_{_{\rm GJ}}|$, assuming that $F_{_{\rm GJ}} \neq 0$. This means that at low 
altitudes the imbalance between $F$ and $F_{_{\rm GJ }}$ (the flux of fictitious charges) that gives 
rise to the accelerating electric field in that region is much smaller than the local GJ flux, and 
therefore there will be no disruption of the steady-state regime of particle flow within the SG. 

\noindent 
At large distances, $1 \ll \eta \ll \eta _{lc}$, 
\be 
|\Delta F| \approx |\Delta F _{\ast } (\eta , \xi ) | \approx 
|F_{_{\rm GJ, 1}}|, 
\label{DeltaF4}
\ee 
where $|F_{_{\rm GJ,1}}(\eta \gg 1)| \gg |F_{_{\rm GJ}}(z=\eta -1\ll 1)|$, 
and $F_{_{\rm GJ}}(z\ll 1)$ is a function of $\xi $ only (i.e. is nearly a constant along the field lines). 

\noindent 
Thus, for low altitudes, up to approximately one-two stellar radii above the PC surface, 
\be 
F_{_{\rm GJ}} \approx F_{_{\rm GJ,0}},
\label{FGJ3}
\ee
and 
\be
|\Delta F | \leq |F_{_{\rm GJ }}|. 
\label{DeltaF5}
\ee
Expressions (\ref{FGJ3}), (\ref{DeltaF5}) imply that the SG boundaries can easily acquire the necessary 
surface charge by means of redistribution of space charges along the field lines in the vicinity of the SG 
boundaries, thus enabling the fulfillment of both the equipotentiality of SG boundaries and continuity of 
${\bf E}_{\perp }$ across them. Note that, the component ${\bf E}_{\perp }$ and therefore the magnitude of 
induced surface charge is mostly determined by $F_{_{\rm GJ,1}}$ (or by $\rho _{_{\rm GJ,1}}$). Since in 
this region $|F_{_{\rm GJ,1}}| \leq |F_{_{\rm GJ,0}}|$, the required surface charge can be easily built up 
by establishing a weak current along the boundary field lines, without violating the GJ condition that 
in a steady-state regime the flux of charges from (to) the stellar surface should be limited by 
$F_{_{\rm GJ,0}}$. We must also note that the presence of a weak current along the SG boundary determined by 
flux $F_{_{\rm GJ,1}}$ is perfectly compatible with equipotentiality of the boundary. For example, this can 
be achieved by establishing a slightly non-homogeneous distribution of charges in a tiny skin layer along 
the boundary, at $\Phi = 0$. 

\noindent 
As we move up to higher altitudes along the SG boundaries where 
\be
|F_{_{\rm GJ,1}}| \geq |F_{_{\rm GJ,0}}|,
\label{FGJ1>FGJ0}
\ee
and where ${\bf E'}_{\perp }$ is mostly determined by the term 
$\sim ({\bf \Omega }\times {\bf r}/c)\times {\bf B}$, the situation changes dramatically.
In this case the equipotentiality of the SG boundaries becomes fundamentally incompatible with a steady state 
regime. In other words, it is very unlikely  that the SG boundaries can be steadily maintained in dynamic 
equilibrium in the presence of a strong ${\bf E'}_{\perp }$ component. Rather, it is this region where the 
centrifugal-slingshot effect makes the electrons/positrons `slip' from the magnetic field lines, and 
therefore effectively prevents charges from building up the otherwise required surplus surface charge on SG boundaries. 
Thus, in a steady-state situation the SG boundaries can be maintained as equipotential, if the approximate 
condition ${\bf E'}_{\perp} = 0$ (or its classical counterpart [\ref{Eperp}], as it will be referred to in the rest 
of the paper) is achieved throughout the SG outermost section. This means that in the outermost section 
of SG the value of $\Delta F (\eta , \xi )$ (see [\ref{DeltaF2}]) cannot grow (because of cross-field motion of 
charges at very high altitudes that effectively destroys the excessive GJ space charge in this region), and it is 
likely to nearly saturate at $\eta \approx \eta _c$ remaining constant along the stream lines. In this case the scalar 
potential $\Phi$ is described by the following Poisson's equation  
\be
{\bf \nabla }\cdot \left( {1\over \alpha } {\bf \nabla }\Phi \right) 
= 4\pi \Delta \rho _{\rm eff}, 
\label{Poisson2}
\ee
where $\Delta \rho _{\rm eff }$ is now determined by $\Delta F(\eta _c, \xi )$ (see formulae [\ref{rhoGJ_2}], 
[\ref{rho}], [\ref{DeltaF1}] and  [\ref{DeltaF2}]), and reads
\begin{eqnarray}
& &\Delta \rho _{\rm eff} = - \rho _0 {{f(\eta )}\over {f(1)}}
{1\over {\alpha \eta ^3}} \{[ 1-\beta + a_1(\eta _c, \xi ) - a_1(1,\xi)\beta ] \cos \chi + 
[b_1(\eta _c, \xi )- b_1(1,\xi)\beta ] \sin \chi \cos \phi _{\rm pc} \} = \nonumber \\
& & - \rho _0 {{f(\eta )}\over {f(1)}}
{1\over {\alpha \eta ^3}} \left\{ \left[\kappa \left( \beta - {1\over {\eta _c^3}} \right) + 1 - \beta \right]
\cos \chi + {3\over 2} H(1)\theta_0 \left( {{H(\eta _c)}\over {H(1)}}\sqrt{{\eta _c f(1)}\over{f(\eta _c)}}-
\beta \right) \sin \chi \cos \phi _{\rm pc} \right\},  
\label{Delta-rho}
\end{eqnarray}
where $\beta $ is defined after eq. (\ref{B}). The parameter $\eta _c$ entering the r.h.s of 
eq. (\ref{Delta-rho}) and defining the altitude of saturation is to be determined through  matching the 
solution at low altitudes of MH03 (see Section 3.2).  

\section{Extended-SG Electrodynamic Solution}

The general formula for the dipolar magnetic field line within the SG can be written as (see MH03)  
\begin{eqnarray}
\sin \theta & = & \sqrt{\eta {{f(1)}\over {f(\eta )}}} \sin \left[ \theta _{_{0,{\rm SG}}} 
\left( 1 \mp {1\over 2} \Delta \xi _{_{\rm SG}} \xi _{\ast } \right) \right] 
\approx \nonumber \\
& & \sqrt{\eta {{f(1)}\over {f(\eta )}}} \theta _{_{0,{\rm SG}}} 
\left( 1 \mp {1\over 2} \Delta \xi _{_{\rm SG}} \xi _{\ast } \right) , 
\label{sin_theta}
\end{eqnarray}
where $\theta _{_{0,{\rm SG}}} = \theta _0 (1- \Delta \xi _{_{SG}}/2) \approx \theta _0 = 
[ \Omega R / c f(1)]^{1/2}$ is the polar angle of the SG central line. This central line separates 
the SG into the innermost half-space and the outermost half-space where the dimensionless colatitude 
of field lines $\xi _{\ast }$ varies from 1 to 0 and from 0 to 1, respectively. Here $\Delta \xi _{_{\rm SG }}$ 
is the latitudinal gap thickness in units of dimensionless colatitude $\xi $ ($\xi = \theta /\theta _0$). Note 
that in a small-angle approximation one may justifiably assume (as it was done in HM03) that 
$\sin \theta \approx \theta $ and $\theta _0^2 \eta \ll 1$.

\noindent
The above formula for the magnetic field lines is based on a canonical definition of the light-cylinder 
radius, $R_{lc} = c/\Omega $. Strictly speaking, this definition is valid only for a nearly aligned rotator 
with poloidal dipolar magnetic field. For a NS with arbitrary obliquity, $\chi $, the footpoints of last open 
field lines slightly deviate from those in the case of an aligned rotator, and the magnetic 
colatitude of the last open field line becomes a function of $\chi $ and $\phi _{\rm pc}$. In addition, 
the rotational distortion of a pure dipolar poloidal magnetic field near the light-cylinder also affects the shape 
of the surface formed by the last open field lines and location of their footpoints. The non-circularity of the PC 
boundary caused by pulsar obliquity and rotation were studied by many authors (see e.g. Cheng, 
Ruderman \& Zhang, 2000 and Dyks, Harding \& Rudak, 2004, for most recent assessment of this issue; and references therein). It is interesting, 
that for $\chi \lsim (60^{\circ}-70^{\circ})$ the assumption that the light-cylinder radius is simply defined as $c/\Omega $ 
seems to be very satisfactory in most theoretical studies on pulsar magnetospheres. For example, for 
$\chi \lsim (60^{\circ}-70^{\circ})$ the effects of pulsar obliquity and magnetosphere rotation discussed in Cheng, Ruderman 
\& Zhang (2000), may distort the circularity of an ``ideal" (very small obliquity and no rotational distortion) 
PC boundary by less than 20-25$\% $.     

We shall reasonably assume that $\Delta \xi _{_{\rm SG}}\ll 1$ (thin SG approximation). For 
$1\ll \eta ^2 \ll \eta _{lc}^2$ (where $\eta _{lc} = R_{lc}/R$), 
the operator $\nabla ^2$ in the l.h.s. of Poisson's equation (see eqn. [\ref{Poisson2}]) reduces to 
\begin{eqnarray}
\nabla _{\Omega }^2 & = & {1\over {\sin \theta }} {\partial \over {\partial \theta }} 
\left( \sin \theta {\partial \over {\partial \theta }} \right) + 
{1\over {\sin ^2 \theta }} {{\partial ^2}\over {\partial \phi _{pc}}^2}\approx 
\nonumber \\
& & {{\partial ^2}\over {\partial \theta ^2}}+{{f(\eta )}\over {f(1)}} 
{1\over {\eta \theta _{_{0, {\rm SG}}}}} {{\partial ^2}\over {\partial \phi _{pc}^2}}.
\label{nabla2}
\end{eqnarray}
Changing the variable $\theta $ to $\xi _{\ast }$ (via relationship [\ref{sin_theta}]), we get 
\be
\nabla _{\Omega }^2 = \left( 1 - \eta {f(1) \over f(\eta )} \theta _{_{0, {\rm SG}}}^2 \right) 
{1\over {\nu _{_{\rm SG}}}} {{f(\eta )}\over {f(1)}}
{1\over {\eta \theta _{_{0,{\rm SG}}}^2}} {{\partial ^2}\over {\partial \xi _{\ast }^2}}
\pm {1\over {\sqrt{\nu _{_{\rm SG}}}}} {\partial \over {\partial \xi _{\ast }}} +
{{f(\eta )}\over {f(1)} } {1\over {\eta \theta _{_{0,\rm SG}}^2}} 
{{\partial ^2}\over {\partial \phi _{pc}^2}} ,
\label{nabla2_2}
\ee
where $\nu _{_{\rm SG}} \equiv (1/4) \Delta \xi _{_{\rm SG}}^2$ (see MH03 for more details). 
Thus, by using expression (\ref{nabla2_2}) the Poisson's equation (\ref{Poisson2}) 
can be rewritten as 
\be
\left( {1\over {\nu _{_{\rm SG}}^{\ast }}} {{\partial ^2}\over {\partial \xi _{\ast }^2}}
\pm {f(1)\over f(\eta )} {1\over {\sqrt{\nu _{_{\rm SG}}}}} \eta \theta _{_{0,{\rm SG}}}^2 
{\partial \over {\partial \xi _{\ast }}} + {{\partial ^2}\over {\partial \phi _{pc}^2}}
\right)  \Phi = - 2 \Phi _0 \theta _{_{0,{\rm SG}}}^2 
({\cal A}\cos \chi + {\cal B} \sin \chi \cos \phi _{pc}), 
\label{poisson3}
\ee
where $\Phi _0 \equiv (\Omega R/c)B_0R$, and 
\be
\nu _{_{\rm SG}}^{\ast } = {{\nu _{_{\rm SG}}}\over {1-\eta f(1)\theta _{_{0,{\rm SG}}}^2/f(\eta )}}. 
\label{nu*}
\ee
Here ${\cal A}$ and ${\cal B}$, according to eq. (\ref{Delta-rho}), are given by 
\be
{\cal A}=\kappa \left( \beta - {1\over {\eta _c^3}} \right) + 1 - \beta ~~~{\rm and }~~~{\cal B}= {3\over 2} 
\theta _0 H(1) \left( {{H(\eta _c)}\over {H(1)}} \sqrt{{{\eta _cf(1)}\over {f(\eta _c)}}} - \beta \right),
\label{AB}
\ee
respectively. In a small-angle approximation (see MH03) one would neglect the second term in the l.h.s. of 
eq. (\ref{poisson3}) and replace $\nu _{_{\rm SG}}^{\ast }$ by $\nu _{_{\rm SG}}$.  

\noindent
We shall search for general solution of eq. (\ref{poisson3}) in the form 
\be
\Phi = \Phi _1 \cos \chi + \Phi _2 \sin \chi \cos \phi _{pc}.
\label{Phi1+Phi2}
\ee
The general solutions for $\Phi _1$ and $\Phi _2$ read 
\begin{eqnarray}
\Phi _1 & = & \pm {2\over \eta } {f(\eta ) \over f(1)} \sqrt{\nu _{_{\rm SG}}}\Phi _0 
{\cal A} (1-\xi _{\ast }) + \nonumber \\
& & 2\left( {f(\eta )\over f(1)} {1\over {\eta \theta _{_{0,{\rm SG}}}}}\right) ^2 
\left( 1 - \eta {f(1)\over f(\eta )} \theta _{_{0,{\rm SG}}}^2 \right) \Phi _0 {\cal A}
\left[\exp{(\mp s)}-\exp{(\mp s\xi _{\ast })} \right],
\label{Phi1}
\end{eqnarray}
and 
\be
\Phi _2 = 2 \Phi _0 \theta _{_{0, {\rm SG}}}^2 {\cal B} \left[ 1- 
{{\cosh(\sqrt{\nu _{_{\rm SG}}}\xi _{\ast })}\over {\cosh(\sqrt{\nu _{_{\rm SG}}})}}  \right] ,
\label{Phi2}
\ee
respectively.

\noindent 
In expression (\ref{Phi1}), 
\be
s = \eta \theta _{_{0, {\rm SG}}}^2 {f(1)\over f(\eta )}
{{\sqrt{\nu _{_{\rm SG}}}}\over {1 - \eta f(1) \theta _{_{0,{\rm SG}}}^2/f(\eta ) }}.
\label{s}
\ee
When $s \ll 1$ (or $1-\eta \theta _0^2 = 1-\eta /\eta _{lc} \gg \sqrt{\nu _{_{\rm SG}}}$),
\be
\Phi _1 \approx \Phi _0 \nu _{_{\rm SG}}^{\ast }\theta _{_{0,{\rm SG}}}^2~{\cal A} (1-\xi _{\ast }^2).
\label{Phi1-2}
\ee
Thus, for $1 - \eta /\eta _{lc} \gg \sqrt{\nu _{_{\rm SG}}}$, the above solutions 
for $\Phi _1$ and $\Phi _2$ yield 
\be 
\Phi (\eta , \xi _{\ast }) = \Phi _0 \theta _{_{0,{\rm SG}}}^2 \left\{ \nu _{_{\rm SG}}^{\ast } 
{\cal A} (1-\xi _{\ast }^2) \cos \chi 
+ 2{\cal B} \left[ 1 - {{\cosh(\sqrt{\nu _{_{\rm SG}}}\xi _{\ast })}\over {\cosh(\sqrt{\nu _{_{\rm SG}}})}}\right] 
\sin \chi \cos \phi _{pc} \right\} .
\label{Phi}
\ee
For $1\ll \eta < \epsilon \eta _{lc}$ (where $\epsilon \approx 0.1-0.5$), eq. (\ref{Phi}) 
can be rewritten as
\begin{eqnarray}
\Phi & = & \left({{\Omega R}\over {c}}\right) ^2 {{B_0}\over {f(1)}} R \left\{ 
\nu _{_{\rm SG}}^{\ast } \left[\kappa \left( \beta-{1\over {\eta _c^3}}\right) + 1 - \beta \right]
(1-\xi _{\ast }^2) \cos \chi + \right. \nonumber \\
& & \left.  3\theta _0 H(1) \left( {{H(\eta _c)}\over {H(1)}}\sqrt{\eta _c {{f(1)}\over {f(\eta _c)}}} -\beta \right)
\left[ 1 - {{\cosh(\sqrt{\nu _{_{\rm SG}}}\xi _{\ast })}\over {\cosh(\sqrt{\nu _{_{\rm SG}}})}}\right] 
\sin \chi \cos \phi _{pc} \right\}.
\label{Phi-appr2}
\end{eqnarray}
In a thin-SG approximation ($\sqrt{\nu _{_{\rm SG}}} \ll 1$), the second term in eq. (\ref{Phi-appr2}) 
can be simplified to yield 
\begin{eqnarray}
\Phi & = & \left({{\Omega R}\over {c}}\right) ^2 {{B_0}\over {f(1)}} R \nu _{_{\rm SG}}\left\{ 
\left[ \kappa \left( \beta -{1\over {\eta _c^3}}\right) + 1 - \beta \right] \left( 1 + {\eta \over \eta _{lc}}\right) \cos \chi + 
\right. \nonumber \\
& & \left.  {3\over 2} \theta _0 H(1) \left( {{H(\eta _c)}\over {H(1)}}
\sqrt{\eta _c {{f(1)}\over {f(\eta _c)}}}-\beta \right) \sin \chi \cos \phi _{pc} \right\} 
(1-\xi _{\ast }^2)  .
\label{Phi-appr3}
\end{eqnarray}
This solution illustrates also that for each magnetic stream line (each $\xi $) within the SG the scalar potential 
$\Phi $ nearly saturates at high altitudes (but well within the light cylinder). The remarkable 
property of the SG solution is that it incorporates the effect of the SG bending 
($\beta $ and $1+\eta/\eta _{lc}$ terms in formula [\ref{Phi-appr3}]). The first ($\beta $-) effect is owing to the dependence of $\rho $ 
on $\eta $ at high altitudes discussed in Section 2.2. The second effect is due to the fact that, as the altitude 
increases, the divergence of the electric field within the SG tends to decrease because of the straightening of the SG (see the first term in eq. [\ref{poisson3}]). However, since the charge per unit length of 
the SG remains constant (see r.h.s. of eq.[\ref{poisson3}]), this geometrical effect is compensated by slight enhancement 
of potential $\Phi $ at higher altitudes. Finally, solution (\ref{Phi-appr3}) reduces to the low-altitude solution 
derived in MH03 if we neglect in formula (\ref{Phi-appr3}) the term $\eta /\eta _{lc}$ and replace $\eta _c $ by $\eta $. Note that in this case $\Phi = -(2/c) \Delta F (1-\xi _{\ast }^2)$, i.e. the scalar potential $\Phi $ is simply 
proportional to the flux of fictitious charge, $\Delta F$. 

Before we discuss how to determine parameter $\eta _c$ entering solution (\ref{Phi-appr3}) we 
shall consider more closely the regions with the so-called ``favorably" and ``unfavorably" curved field lines 
introduced by Arons \& Scharlemann (1979). The region with favorably curved field lines simply 
corresponds to that with $\cos \phi _{pc} > 0$ (electron acceleration, for normal polarity), 
and the region with unfavorably 
curved field lines corresponds to that with $\cos \phi _{pc} < 0$ (positive charge acceleration). Note 
that for the reversed polarity pulsar (see definition below formula [\ref{f()}]) 
the sign of accelerating particles reverses. Since in the electrodynamic 
model of Arons \& Scharlemann (1979) the accelerating electric field is solely determined by the component 
$\propto \sin \chi \cos \phi _{pc}$, the electrons or positive charges (e.g. positrons) may accelerate in their model 
only along favorably or unfavorably curved field lines, respectively. In this sense, in our model the presence of the 
component $\propto \cos \chi $ (which is also independent of $\phi _{pc}$) tends to make all field lines 
`favorably curved' in that the same sign of charge is accelerated outward over the entire PC. 
However, for very large obliquities and/or very short spin periods the favorably curved 
field lines may again become unfavorable for electron acceleration, because of the negative contribution 
from the term $\propto \sin \chi \cos \phi _{pc}$  (for the region where $\cos \phi _{pc} < 0$). Now consider the situation 
where in the formula for $\Phi $ the component which is $\propto \cos \chi $ dominates. In this 
case the value of $\eta _c$ can be estimated from matching the asymptotic value of $E_{\parallel }$  with the 
value of $E_{\parallel }$ generated at small altitudes, which translates into 
$3\kappa /\eta _c^4 \sim (3+5\kappa )/8\eta _{lc}$. For the parameters of the Crab pulsar and assuming very small 
obliquity this condition would result in $\eta _c \sim 3$. For $\chi = 60^o$ and $\cos \phi _{pc} < 0$, we 
come up with the estimate $\eta _c \sim 1.2-1.3$. 

\noindent
Using formula (\ref{Phi-appr3}) and $E_{\parallel } = - (1/\alpha )\nabla _{\parallel }
\Phi = -(1/R)\partial \Phi /\partial \eta $ we can get, for $\eta \gsim \eta _c$,  
\begin{eqnarray}
E_{\parallel, {\rm high }} &\approx &- {3\over 8}\left({{\Omega R}\over {c}}\right) ^3 {{B_0}\over {f(1)}} 
\nu _{_{\rm SG}} \left\{ \left[ 1+ {1\over 3}\kappa \left( 5 - {8\over {\eta _c^3}}\right) +2{\eta \over \eta _{lc}} \right] \cos \chi + \right. \nonumber \\
&& \left. {3\over 2} \theta _0 H(1) \sin \chi \cos \phi _{pc} \right\}  (1-\xi _{\ast }^2) .
\label{Epar-appr1}
\end{eqnarray}
According to formula (\ref{Epar-appr1}), 
\be
|E_{\parallel,{\rm high }}|\sim {{B_0}\over {\eta _{lc}^3}} \nu _{_{\rm SG}} \ll 
B(\eta \sim \eta _{lc}) \sim |E_{\perp }(\eta _{lc})|,  
\label{Epar-appr2}
\ee
which is the condition accompanying the regime given by Eqn (\ref{Eperp}).

\subsection{Extended Acceleration Within the SG in Curvature-Radiation Reaction Limit}

Let us estimate the characteristic electron Lorentz factor, $\gamma $, in the regime of extended acceleration 
within the SG in the situation where the gain in electron kinetic energy is almost exactly compensated 
by the CR reaction,  i.e. where the condition 
\be
eE_{\parallel, {\rm high }} \approx {2\over 3}{{e^2}\over {\rho _c^2}}\gamma ^4
\label{CRRL}
\ee
is satisfied. Here $\rho _c \sim (4/3)~R~(\eta \eta _{lc})^{1/2}$ is the characteristic curvature radius 
of the field lines in the SG. 

\noindent
By using formula (\ref{Epar-appr1}) in the above equation, we arrive at
\be
\gamma \sim 3 \cdot 10^7 \left[
(2.5+0.6\kappa _{0.15}) B_{12} {{R_6^3} \over P_{0.1}} {\eta \over \eta _{lc}} 
\nu _{_{\rm SG}} (1-\xi _{\ast }^2) |\cos \chi |\right]^{1/4} ,
\label{gamma-CRRL}
\ee
where $\kappa _{0.15} = \kappa /0.15 \approx I_{45}/R_6^3$; $I_{45}=I/10^{45}$ g$\cdot $cm$^2$, $I$ is the 
NS moment of inertia. Here, for the sake of simplicity, we assume that the term $\propto \cos \chi $ 
dominates. 

\noindent
Using parameters of the Crab pulsar, $B_{12} = 8$, $P_{0.1} = 0.33$, $R_6 = 1.6$, and 
for $\nu _{_{\rm SG}} \sim 0.05$ and $\eta \sim 0.5\eta _{lc}$ we 
get 
\be
\gamma \sim 4\cdot 10^7 [(2.5+0.6\kappa _{0.15})(1-\xi _{\ast }^2) |\cos \chi | ]^ {1/4}.
\label{gamma-CRRL2}
\ee
This estimate implies that the energetics of the extended SG region is about the same as that of the region 
near the PC surface, even though the physical conditions for pair production and generation of high-energy 
emission in the outermost part of the SG are quite different (we shall address this issue in the next Section). 

\subsection{Numerical Results}

In this section we will present some results of numerical simulations of the extended SG solution derived 
earlier in Section 3 
for a particular case.  The results shown in this Section are meant to illustrate the most important features of the proposed acceleration model.  A more extensive exploration of pulse profile properties and spectra, and fitting to 
observed high-energy pulsars, will appear in another paper.  We wish to have an approximate expression for the accelerating $E_{\parallel}$ in the SG from the PC surface to near the light cylinder.  As discussed earlier in 
Section 3, we are not able to derive a single solution valid at all altitudes, but were able to find expressions both within a few stellar radii of 
the PC surface and at large radii, when condition (\ref{Eperp}) dominates the solution.  Our first step is thus to find an expression that best matches the solutions in the two regions, by finding the value of parameter $\eta_c$.  We will 
then use this approximate expression to simulate acceleration and pair cascade radiation of primary electrons from the stellar surface. 

\subsubsection{Matching Low Altitude and High Altitude Solutions for $E_{\parallel}$}

Eqn (12) of MH03 provides an expression for the scalar potential $\Phi$ valid in the SG at low altitudes (for radii $\eta \lsim \eta_c$).  The low altitude accelerating 
$E_{\parallel} = -(1/\alpha )\nabla _{\parallel } \Phi$ is thus
\be
E_{\parallel, {\rm low}}  \simeq  -3\left({{\Omega R}\over {c}}\right) ^2 {{B_0}\over {f(1)}} \nu _{_{\rm SG}}
(1-\xi _{\ast }^2)\left[{\kappa\over \eta^4}  \cos \chi + 
 {1\over 2} \theta _0 H(1) \delta(\eta) \sin \chi \cos \phi _{pc} \right], 
\label{Epar - lo}
\ee
where $\delta (\eta ) = d\ln (H\theta)/d\eta $ (MH97) varies between $\sim 0.5$ and 1. 
One can write $\delta (\eta ) = 0.5 - d\ln H/d\ln x \approx 0.5+0.25x+0.39x^2$ [where $x= \varepsilon /\eta $; 
see also eqs (\ref{f()}) and (\ref{H})], which, for a canonical NS, gives $\delta (1) \approx 0.7$.  As noted
earlier, $H(1) \approx 0.8$ for a canonical NS.

At high altitude, $\eta \gsim \eta_c$, we have argued that the effective flux, $\Delta F$, will nearly saturate 
and that the $E_{\parallel,{\rm high }}$ will be given by expression (\ref{Epar-appr1}). In order to match these two solutions 
at high and low altitude, we write a general expression of the form
\be
E_{\parallel } \simeq E_{\parallel, {\rm low}}\exp{[-(\eta-1)/(\eta_c-1)]} 
\label{Epar - tot} + E_{\parallel, {\rm high }} 
\ee
and determine $\eta_c$ to give a smooth transition between $E_{\parallel, {\rm high}}$ and $E_{\parallel, {\rm low}}$ 
given by expressions (\ref{Epar-appr1}) and (\ref{Epar - lo}), respectively. Formula 
(\ref{Epar - tot}) is valid (both for favorably and unfavorably curved field lines) when 
$[\kappa \cos \chi + (1/2)\theta _0 H(1)\delta (1)\sin \chi \cos \phi _{pc}] > 0$ or, more precisely, 
when $\Delta F(\eta )/F_0 > 0$ (for $\eta \lsim \eta _c$). For a reversed polarity pulsar, formula (\ref{Epar - tot}) 
is also applicable for electron acceleration for nearly orthogonal rotator and for unfavorably curved 
field lines.  

For nearly aligned ($0^{\circ} \lsim \chi \lsim 60^{\circ}$) or nearly anti-aligned ($120^{\circ} \lsim \chi \lsim 180^{\circ}$) cases, there will be continuous acceleration of electrons from the PC surface up to very high altitudes
on both favorably ($\cos\phi_{pc} > 0$) and unfavorably ($\cos\phi_{pc} < 0$) curved field lines.  For nearly orthogonal rotators
($80^{\circ} \lsim \chi \lsim 100^{\circ}$), there will continuous acceleration of electrons/positrons from the PC 
surface up to very high altitudes on favorably 
curved field lines for normal/reversed polarity, and continuous acceleration of positrons/electrons on unfavorably curved
field lines for normal/reversed polarity.  At intermediate inclination ($60^{\circ} \lsim \chi \lsim 80^{\circ}$ and
$100^{\circ} \lsim \chi \lsim 120^{\circ}$), there will be continuous acceleration of electrons on favorably/unfavorably curved
field lines for normal/reversed polarity.  However, on unfavorably/favorably curved field lines for normal/reversed
polarity, there may not be continuous acceleration of electrons from the surface, since the first term ($\propto \cos \chi $) in eq. (\ref{Epar - lo}) dominates only near the NS surface and the second term accelerates positrons.  At higher
altitudes, where eq. (\ref{Epar-appr1}) applies, there will again be a regime favorable for electron acceleration.  
However, this situation is unstable since it implies a build-up of charge.  The system 
will likely become self-limiting, possibly shutting down all charge flow along this field line.
This situation may be favorable for the operation of an outer gap on such field lines 
(see e.g. Cheng, Ho \& Ruderman 1986). 
Thus, the extended SG may not provide steady-state particle acceleration at all inclinations. 
The value for $\eta_c$ will vary with inclination angle and may approach e.g. close to the stellar surface for 
$\chi \gsim 60^{\circ}$. 

We have chosen for our numerical simulations an inclination angle $\chi= 45^{\circ}$ in 
order to illustrate the extended SG solution we discuss in this paper. For $\chi= 45^{\circ}$ we estimate that 
$\eta_c = 1.4$.

\subsubsection{Geometry of Acceleration and Emission in the Extended SG} 

Using the approximate expression for $E_{\parallel}$ in eq. (\ref{Epar - tot}), we have simulated the acceleration and cascade radiation of a primary electron beginning at the PC surface and extending to $0.8 R_{lc}$.  The pair cascade simulation code that we use is based on one developed and described in several previous papers (Daugherty \& Harding 1996, Harding et al. 1997, MH03).  The code follows a primary electron along a number of magnetic field lines above the PC, starting at rest and tracking its Lorentz factor from a given accelerating $E_{\parallel}$ and energy losses due to CR and inverse Compton scattering.  The paths of photons from curvature and inverse Compton radiation are traced to determine if they pair produce (by the one-photon process) or escape, at which point they are accumulated in a 3-dimensional array of energy and angle.  Photon splitting, as treated in Harding et al. (1997) has not been included in the present calculation, and indeed is not important for most pulsars in the SG at high altitudes where the magnetic field is well below the critical value.  Synchrotron radiation from several generations of secondary pairs is also traced, until all radiated photons escape the magnetosphere. As in MH03, we adapt the primary acceleration part of the code to include only acceleration in the SG; thus we inject primary electrons only along the last open field lines at equally spaced azimuth about the magnetic pole.  We have assumed a static vacuum dipole model for the magnetic field and thus do not consider the distortions to the PC footpoints or the bending-back of field lines near the light cylinder that appear in retarded vacuum dipole models (e.g. Dyks et al. 2003).  Since the pair cascades occur at relatively high altitude, above the pair formation front forming the inner bounary of the SG, where the $E_{\parallel}$ is screened, the secondary pairs are not assumed to accelerate.  With the $E_{\parallel}$ of eq. (\ref{Epar - tot}), only primary electrons continue accelerating to high altitude within the SG.  As shown in Section 3.1, the accelerating primary Lorentz factors become radiation-reaction limited by curvature losses at altitudes within and above the pair cascade zone $(1.5 - 3)~R$.  Thus, these particles will continue radiating well above the cascade zone with Lorentz factor given approximately by eq. (\ref{gamma-CRRL}).

Figure 1 shows an observer angle-phase ($\zeta-\phi$) plot of escaping photons above 100 MeV for pulsar inclination angle $\chi= 45^{\circ}$ and parameters $P = 0.033$ s and $B_0 = 8 \times 10^{12}$ G of the Crab pulsar.   One can see the pattern of field lines due to the discrete spacing of azimuthal injection points of the primary electrons.  The intensity of radiation along each field line is shown as a grey scale.  One of the important features to note is the ring of enhanced emission around each magnetic pole, which is the hollow emission cone formed by the pair cascade radiation.  This component terminates at altitudes around $(3-4)~R$.  Another important feature are the lines of caustic emission trailing out beyond the pair cascade rings from each pole.  The caustic emission comes from electrons radiating curvature radiation between altitudes of $(0.1 - 0.7)~R_{lc}$ along the trailing field lines.  As first noted by Morini (1983), who considered only the caustic emission from  one pole, and recently  by Dyks \& Rudak (2003), who considered emission from both poles, the caustics form because the positive phase delays from different altitudes along the trailing dipole field lines are nearly completely cancelled by negative phase shifts caused by aberration and time-of-flight.  Since our code includes such relativistic effects, the two-pole caustic emission seen by Dyks \& Rudak (2003) is a feature of extended SG acceleration.  

Pulse profiles result from slicing the phase plot of Figure 1 at constant observer angle, $\zeta$.  It is apparent that observers having $30^{\circ} \lsim \zeta \lsim 70^{\circ}$ and $120^{\circ} \lsim \zeta \lsim 160^{\circ}$ will see pair cascade emission, in which case either a single peak or two peaks with separation dependent on observer angle and inclination result.  Observers having $60^{\circ} \lsim \zeta \lsim 120^{\circ}$ will cross lines of caustic emission from both poles, resulting in two narrow peaks in the pulse profile.  Since the caustic emission lines are slanted with respect to each other in the phase plot, the peak separation will vary with observer angle and will, in general, not be $180^{\circ}$. Figure 2 shows a profile for $\zeta = 113^{\circ}$, in which there are two peaks separated by about $140^{\circ}$ with a higher level of emission between the peaks, about what is measured for the separation of the peaks in the Crab and Vela high-energy pulse profiles.  Note that the closest approach to one of the magnetic poles occurs at phase $-180^{\circ}$, where an observer could also see the edge of a radio conal emission component.  This could explain the position of the precursor in the Crab pulsar, or the phase of the radio pulse in the Vela pulsar, both of which lead the double $\gamma$-ray peaks.  An identical profile would be seen at observer angle $67^{\circ}$, which is approximately the viewing angle inferred from the Chandra image of the inner Crab nebula in X-rays (Weisskopf et al. 2000).  The caustic emission occurs on the unfavorably curved field lines, i.e. those curving away from the rotation axis at observer angles $\chi < \zeta < 180^{\circ} - \chi$.  The inclusion of the frame-dragging component of the potential, which is the only one that survives at high altitude and is not dependent on magnetic azimuth, is thus critical to allowing caustic emission.  At certain (very limited) observer angles, it is possible to view both cascade and caustic emission, in  which case there would be three or four peaks in the profile of varying relative intensity.  

Acceleration and cascade emission along field lines interior to the SG (`core' 
emission) is also
expected, but has not been included here.  Since the pair cascades on the interior field lines
occur much closer to the NS surface, the emission would form a hollow cones of smaller
radius in the phase plot.  For observer angles crossing near the magnetic pole, such core emission
would produce flux in the interpeak region of the profile.  For observer angles crossing the
caustics, the core emission would not produce much additional flux in the profile, at least
at high energies, since the electric field is screened at low altitude on the interior field
lines and primaries do not continue to accelerate above the cascade region as they 
do in the SG.

In the present simulation, we have included only CR of the primary electrons accelerating at high altitudes, which forms the caustic emission.  As noted earlier, they have reached radiation-reaction limit so that their CR spectrum will be quite hard (photon index $-2/3$).  This spectrum is much harder than the observed spectra of $\gamma$-ray pulsars.  However, the primary electrons will also be scattering sources of soft photons, such as radio, optical and infrared.  We are investigating the possibility that inverse-Compton scattering of radio emission along the same field lines, forming the conal component in phase with the high-energy peaks, may account for the spectrum of the Crab pulsar.

\section{Discussion and Conclusions}

In this paper we extend our previous study (MH03) of pulsar SGs to investigate the regime of acceleration 
of primary electrons and high-energy emission to very high altitudes.  Incorporating the effect of cross-field 
motion of charges near the light cylinder on the distribution of accelerating potential within the entire SG,
we derived the explicit expressions 
for the accelerating electric field in the space-charge-limited flow approximation.  We have modeled the particle acceleration up to high altitudes and generation of high-energy emission within the SG, illustrating the resulting high-energy radiation pattern and light curves for the case of the Crab pulsar.  The primary (space-charge-limited) flow within the SG becomes radiation-reaction limited at high altitudes, such that the energy gain from acceleration is nearly compensated by the CR losses.  The resulting emission pattern exhibits both the hollow cones centered on each magnetic 
pole  
(the corresponding phenomenological model was first discussed in connection with radio emission by Komesaroff 1970, Radhakrishnan 1969, and 
Radhakrishnan \& Cooke 1969), due to pair cascades on the inner edge of the SG, as 
well as the caustics along the trailing field lines at high altitude noted in previous studies (Morini 1983, Romani \& Yadigaroglu 1995, Cheng et al. 2000, Dyks \& Rudak 2003).  The extended SG acceleration allows the caustic emission to be viewed from both magnetic poles simultaneously, naturally producing the widely-spaced double-peaked profiles seen in many high-energy pulsars.
      
Note that in a vacuum NS magnetosphere with the strong magnetic field prohibiting transfield motion, the GJ 
space charge may serve as a source of electric field accelerating test particles (e.g. electrons/positrons) along the 
magnetic field.  In the SG geometry with space-charge-limited flow (nearly GJ near the surface), the occurrence of a 
steady-state cross-field drifting at very high altitudes tends to reduce the GJ space charge (see condition 
[\ref{Eperp}] and related discussion) within the SG, maintaining the SG boundaries as equipotential surfaces. 
The reduction of the GJ space charge is accompanied by a steady-state distribution of the effective flux, $\Delta F$, 
within the SG. Our model requires that the source of a steady-state accelerating electric field within the SG 
establishes at relatively low altitudes (e.g. over a few stellar radii above the PC) and remains constant along 
the magnetic stream lines up to very high altitudes.

Whether or not a steady-state particle flow can be achieved on a given field line will depend on the relative 
importance of the terms $\propto \cos \chi$ and $\propto \theta_0 \sin \chi \cos \phi _{pc}$ near the NS surface, which depends on the pulsar inclination angle $\chi$, the magnetic azimuth $\cos \phi _{pc}$ and the PC opening angle $\theta_0$.   
If the term $\propto \cos \chi$ dominates, then $E_{\parallel}$ accelerates the same sign of charge at all altitudes and steady-state flow can be maintained.  However, if the term  $\propto \theta_0 \sin \chi \cos \phi _{pc}$ begins to dominate near the surface for large $\chi$ on unfavorably curved field lines ($\cos \phi _{pc} < 0$), then $E_{\parallel}$ changes sign and may prevent a steady-state flow of charge (or any charge flow) along that field line 
(see Section 3.2.1).

Even if $E_{\parallel}$ does not change sign above the PC, the possibility we have explored in this paper of a steady-state regime of acceleration of primaries from the PC surface up to the light cylinder implicitly assumes a free supply of charge from the NS.  We do not exclude a scenario with non-stationary or insufficient supply of 
primaries from the PC surface, in which case there could be a regime of intermittent charge flow.  
Note that we have briefly discussed a similar 
possibility in the case of acceleration of primaries over the main part of the PC (see HM01).  

Outer gap models (e.g. Cheng et al. 1986, Romani 1996), that assume vacuum gaps form along field lines above the null
charge surface, require little or no current flow into the gaps from outside.  In the model presented in this paper, steady current flow can exist from the PC to the light cylinder along even some unfavorably curved field 
lines where outer gaps are assumed to exist.  In this case the space-charge limited current flow of nearly GJ surface value, combined with cross-field particle motion, would prevent the formation of an outer gap.  In the case of very high inclination angle, where either non-steady current flow or insufficient charge supply from the surface exists, outer gaps may still form (see Section 3.2.1). In this case, however, the radiation which forms caustic emission above the null 
surface can be viewed from only one pole (Romani \& Yadigaroglu 1995, Cheng et al. 2000).  The geometry of the extended 
SG emission is thus clearly distinct from that of the outer gap emission and will have observable consequences, as has 
been discussed by Dyks \& Rudak (2003) and Dyks et al. (2003) and will be detailed in future publications.

\acknowledgments 
We would like to thank Vladimir Usov, Kouichi Hirotani, Bing Zhang and Jarek Dyks for valuable discussions
and comments on the manuscipt.  We also acknowledge support from the NASA Astrophysics Theory Program.

\newpage

~
\figureout{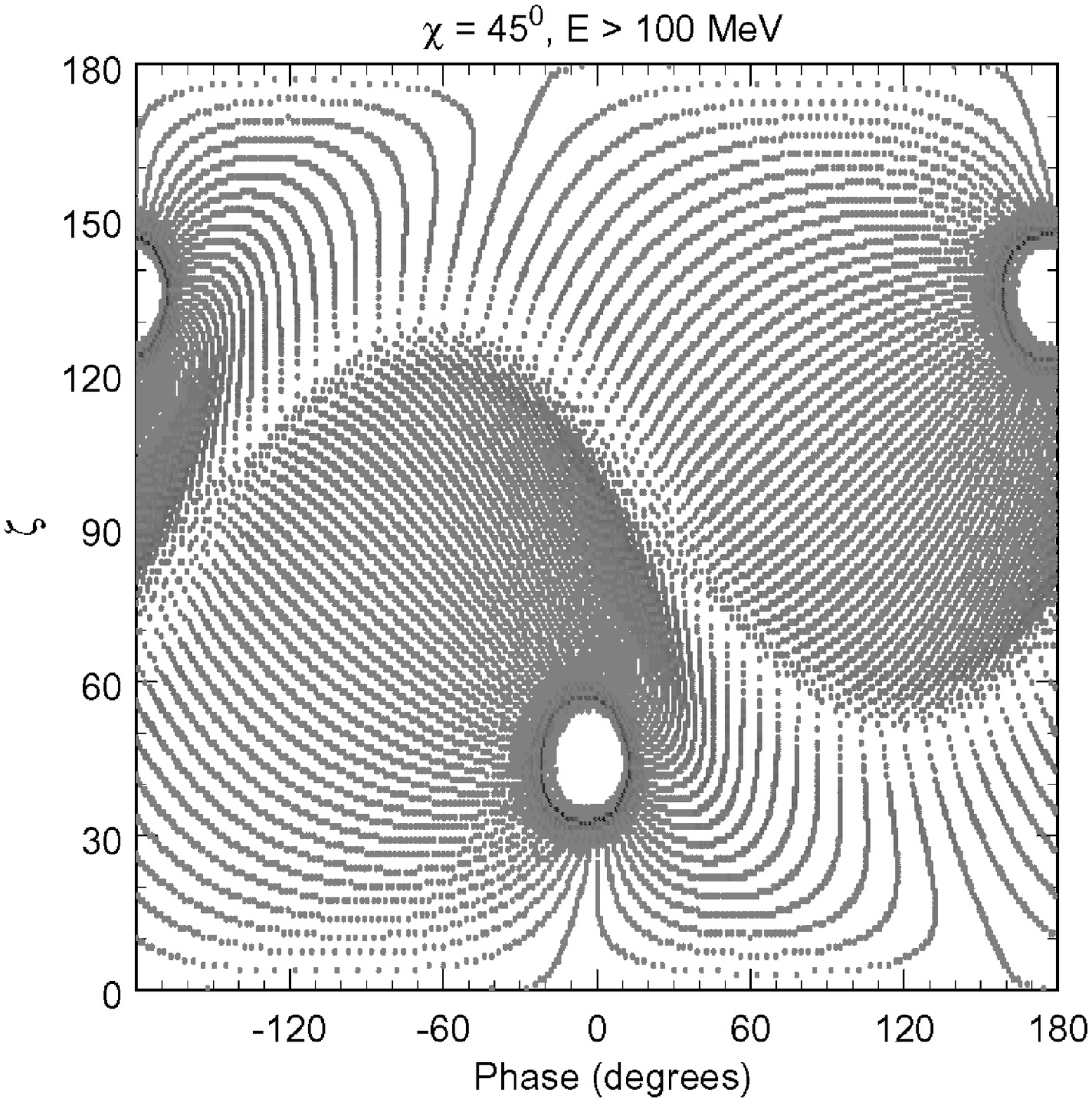}{0}{
Map of emission above 100 MeV as a function of phase $\phi$ and observer angle $\zeta$ with respect to the
rotation axis for a pulsar with magnetic inclination angle of $\chi = 45^{\circ}$, period $P = 33$ ms and
surface magnetic field $B_0 = 8 \times 10^{12}$ G.  Magnetic poles are located at ($\phi = 0^{\circ}, 
\zeta = 45^{\circ}$) and ($\phi = 180^{\circ}, \zeta = 135^{\circ}$).   
    }    

\figureout{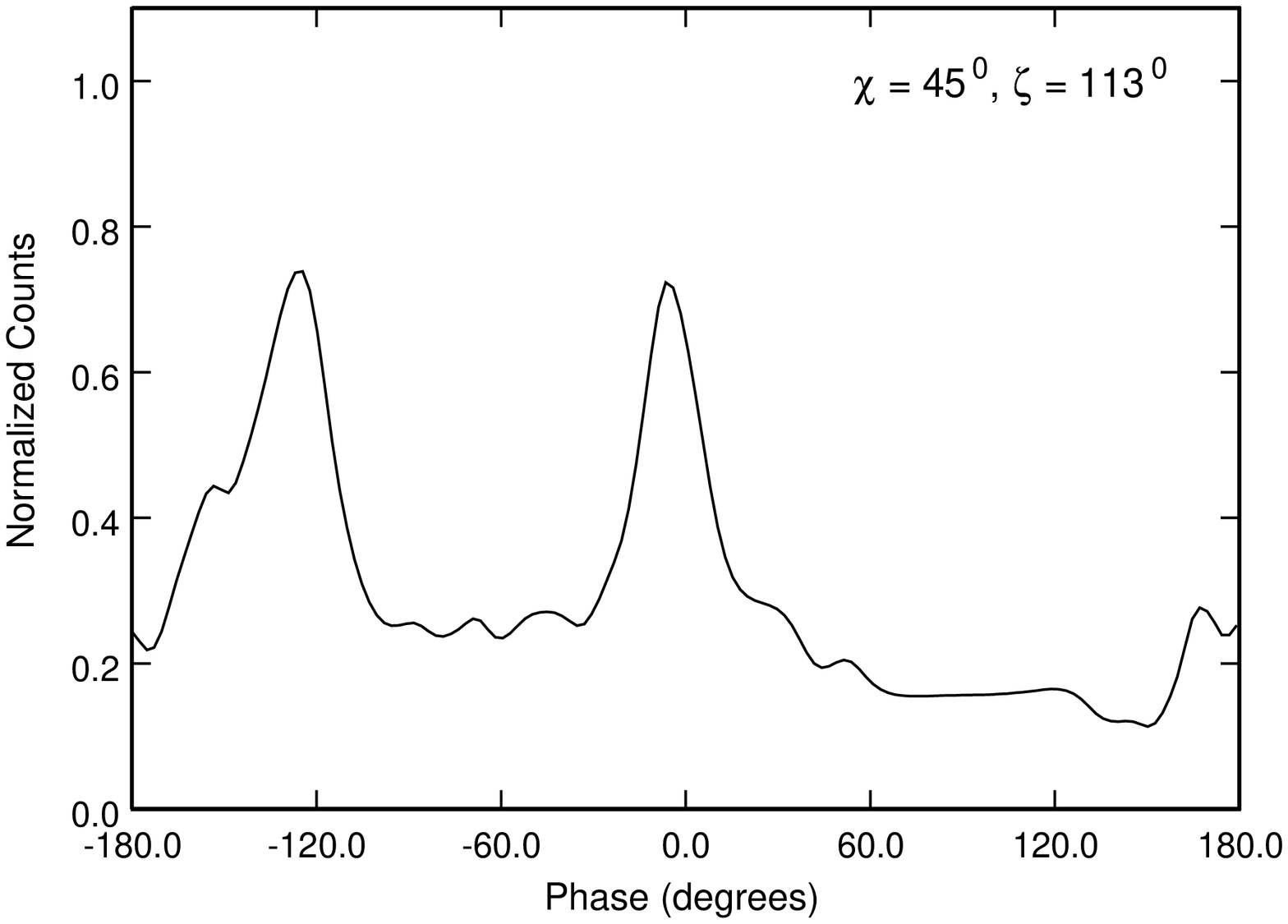}{0}{
Pulse profile for emission above 100 MeV for the case of Figure 1 and observer angle $\zeta = 113^{\circ}$.
    }    

\end{document}